# Tap Density Equations of Granular Powders Based on the Rate Process Theory and the Free Volume Concept

Tian Hao

*Nutrilite Health Institute*
*5600 Beach Blvd., Buena Park, CA 90621, USA*

Tap density of a granular powder is often linked to the flowability via Carr Index that measures how tight a powder can be packed, under an assumption that more easily packed powders usually flow poorly. Understanding how particles are packed is important for revealing why a powder flows better than others. There are two types of empirical equations that were proposed to fit the experimental data of packing fractions vs. numbers of taps in literature: The inverse logarithmic and the stretched exponential. Using the rate process theory and the free volume concept, we obtain the tap density equations and they can be reducible to the two empirical equations currently widely used in literature. Our equations could potentially fit experimental data better with an additional adjustable parameter. The tapping amplitude and frequency, the weight of the granular materials, and the environment temperature are grouped into one parameter that weighs the pace of packing process. The current results, in conjunction with our previous findings, may imply that both "dry" (granular) and "wet" (colloidal and polymeric) particle systems are governed by the same physical mechanisms in term of the role of the free volume and how particles behave (a rate controlled process).

Granular powders have been widely used in many practical areas spanning from pharmaceutical, nutritional, food, and engineering industries. Tap density of a granular powder is often linked to the flowability via Carr Index that measures how tight a powder can be packed, under an assumption that more easily packed powders usually flow poorly. Understanding how particles are packed is important for revealing why a powder flows better than others and how we can handle powders properly during powder processing and manufacturing. The bulk density of granular powders has been found to change with the number of taps applied to whole powder systems. There are many works addressing granular packing with a variety of tapping methods (Knight, et.al.1995, Vandewalle, et.al. 2007). A heuristic logarithmic law is proposed by Knight (Knight, et.al. 1995) based on the tapping experimental data of mono-dispersed 2 mm glass beads and shown below:

$$\phi = \phi_m - \frac{\phi_m - \phi_0}{1 + B\ln\left(1 + \frac{n}{\tau}\right)} \quad (1)$$

Where $\phi_0$ is the initial particle volume fraction, $\phi_m$ is the maximum packing fraction, and $B$ and $\tau$ are two constant dependent on the tapping amplitude, frequency, size of tube, particle sizes and shape, etc.. Eq. (1) is frequently called Chicago inverse logarithmic law. Using the free volume concept, Boutreux and de Gennes theoretically developed a tap density equation very similar to Eq. (1) (Boutreux and de Gennes 1997). They assumed that the particle fills into the free volume available for an individual particle during tapping process. The possibility of any free volume per particle larger than the individual particle volume $V_{ip} = \frac{4\pi r^3}{3}$ may be expressed with the Poisson's distribution and the final equation they obtained is:

$$\phi = \phi_m - \frac{\phi_m^2}{\ln n + \ln n_m} \quad (2)$$

Where $n_m$ is a constant that roughly tells us how many taps needed to reach a steady-state tap density. Several researchers indicated that the inverse logarithmic law may fit the data at the beginning of tapping process but fail to correctly fit the data at the final tapping process up to the steady-state plateau (Philippe and Bideau 2002, Vandewalle et.al 2007). A stretched exponential law that is often used as a phenomenological description of relaxation in disordered systems is found to better fit the experimental data (Philippe and Bideau 2002) with the following form:

$$\phi = \phi_m - (\phi_m - \phi_0)\exp\left[-\left(\frac{n}{\tau}\right)^\beta\right] \quad (3)$$

Where $\beta$ is a number of order 1, the stretching of the exponential. Knight (Knight et. al. 1995) found that their data can be fitted with Eq. (3), too, though the inverse logarithmic law gives a better fit. However, Vandewalle (Vandewalle, et.al., 2007) pointed out that both laws have limitations if they were used to fit their data. Therefore, there is a need for new equations that may work better for granular powder tapping process.

In this article, we will use the free volume concept and rate process theory with additional assumptions to derive tap density equations. We will address the inverse logarithmic law first and the stretched exponential law afterwards. The comparisons between newly derived equations and the currently widely used empirical equations are provided.

Let's consider a simple granular system with initial particle volume fraction $\phi_0$, volume $V_s$, and maximum packing fraction $\phi_m$. After the granular system is tapped for $n$ times, the particle volume fraction is changed from $\phi_0$ to $\phi$. Since both $\phi_0$ to $\phi$ are smaller than $\phi_m$, there is free volume existing in the system and unoccupied by particles. The free volume in a system of particle volume fraction $\phi$ may be expressed as:

$$V_f = V_s(\phi_m - \phi) \quad (4)$$

At the beginning of a tapping process, the initial particle volume fraction is $\phi_0$, thus the free volume may be expressed as

$$V_{f0} = V_s(\phi_m - \phi_0) \quad (5)$$

We may consider the tapping process as a rate process (Glasstone, et.al, 1941) and treat the number of tapping $n$ as a continuous variable. The tapping process rate constant $k$ may be related to how the free volume of a granular system decreases with the number of taps, therefore one may assume:

$$k = A\frac{V_f}{V_{f0}} = \frac{A(\phi_m - \phi)}{\phi_m - \phi_0} \qquad (6)$$

Where *A* is a constant, dependent on the tapping amplitude and frequency, the weight of powders, particle size distribution and so on. We may further assume that the change of particle volume fraction with the tap number *n* is directly proportional to the available free volume per unit volume in the system, as the more free volume the system has, the faster the particle volume fraction will increase. In contrast, we know that the particle volume fraction increase will be much slower at the end of the tapping process when the tap density approaches to a plateau region. Therefore, the change of particle volume fraction with the number of taps should be inversely proportional to the number of taps, with a fast pace at low tap numbers but a slow pace at high tap numbers. Suppose that for reaching a steady-state tap density, the maximum number of taps, $n_m$, is required to make the particle volume fraction very close to $\phi_m$. Thus $(n/n_m)$ may be used to scale how close the number of taps is, in comparison with the final tap numbers, $n_m$. So the change of particle volume fraction with the number of taps should inversely proportional to $(n/n_m)$ instead of the number of taps, n. With those assumptions, one may easily write:

$$\frac{d\phi}{dn} = k\frac{(\phi_m - \phi)}{n/n_m} \qquad (7)$$

Note that the free volume per unit volume is used in above equation. Since *n* is treated as a continuous number, for avoiding *n*=0 and Eq. (7) becoming invalid, one may use $(n/n_m + 1)$ to replace $n/n_m$ and re-write Eq. (7) as:

$$\frac{d\phi}{dn} = k\frac{(\phi_m - \phi)}{n/n_m + 1} \qquad (8)$$

Substituting Eq. (6) into Eq. (8) and re-arranging yields

$$\frac{(\phi_m - \phi_0)}{A(\phi_m - \phi)^2} d\phi = \frac{dn}{n/n_m + 1} \qquad (9)$$

Integrating both sides within the volume fraction range ($\phi_0$, $\phi$) and using the boundary condition (*n*=0, the particle volume fraction is $\phi_0$, and *n*=*n*, the particle volume fraction is $\phi$) yields:

$$\frac{(\phi_m - \phi_0)}{(\phi_m - \phi)} - 1 = \frac{A}{n_m} \ln\left(\frac{n}{n_m} + 1\right) \qquad (10)$$

Re-arranging Eq. (10) yields

$$\phi = \phi_m - \frac{\phi_m - \phi_0}{1 + \frac{A}{n_m}\ln\left(1 + \frac{n}{n_m}\right)} \qquad (11)$$

Eq. (11) has an exact form as Eq. (1), clearly telling what are the physical meanings of the constants, *B* and $\tau$, in the empirical logarithm law expressed in Eq. (1).

Hao has developed a precise way for calculating the free volume of particulate systems using the inter-particle spacing (IPS) concept (Hao, 2005, Hao 2006, Hao 2008). The inter-particle spacing (IPS) that scales the distance between two particle surfaces was used for estimating the free volume of whole systems to derive the viscosity of colloidal suspension systems (Hao 2005, Hao 2008). For calculating IPS, Hao (Hao 2006) used Kuwabara's cell model (Kuwabara 1959) that was extended by many other researchers (Kozak and Davis 1986, Levine and Neale 1974, Oshima 1997) for calculating the electrophoretic and electroacoustic mobility of particles. The cell model assumes that each particle is surrounded by a virtual cell (see Figure 1) and the particle/liquid volume ratio in a unit cell is equal to the particle volume fraction throughout the entire system. Given that the particle is spherical and mono-dispersed, the IPS should be zero when particles reach the maximum particle packing fraction, $\phi_m$, as particles intimately contact each another at the maximum packing fraction. When the particle volume fraction, $\phi$, is less than $\phi_m$, there is a free volume unoccupied by particles. If the volume of a system is $V_s$, then the free volume of the particles should have in this system may be expressed as Eq. (4). The free volume per particle should be:

$$V_s (\phi_m - \phi) / (V_s \phi / V_{ip}) = (\phi_m - \phi) V_{ip} / \phi \qquad (12)$$

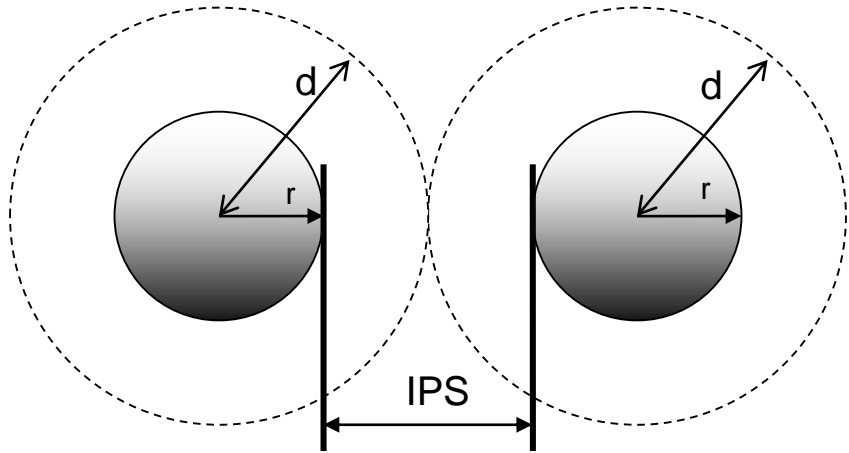

*Figure 1. Illustration of Kuwabara's cell model used for calculating IPS in a particulate system. From Hao, 2005, Electrorheological fluids: the non-aqueous suspensions, Amsterdam: Elsevier. With permission.*

Where $V_{ip}$ is the volume of individual particle, and is equal to $(4\pi r^3)/3$. $r$ is the particle radius. The total volume that each particle occupies in the system is the volume of each individual particle plus the free volume per particle:

$$V_{ip} + (\phi_m - \phi) V_{ip} / \phi = \phi_m V_{ip} / \phi \qquad (13)$$

If the radius of particle plus the virtue cell is $d$, then the IPS defined in Figure 1 may be expressed as:

$$IPS = 2(d - r) \qquad (14)$$

Since $d$ can be calculated from Eq. (13) using the following equation:

$$\frac{4\pi}{3} d^3 = \frac{\phi_m V_{ip}}{\phi} \qquad (15)$$

Eq. (14) is thus rewritten as:

$$IPS = 2(\sqrt[3]{\phi_m / \phi} - 1)r \qquad (\phi \leq \phi_m) \qquad (16)$$

Eq. (16) indicates that IPS is zero when the particle volume fraction reaches the maximum packing fraction, which is consistent with our assumption at the beginning. The parameter, $\phi$, should be always less than $\phi_m$. Once the maximum packing fraction and particle size of a powder system is known, the IPS can be easily estimated using Eq. (16). Suppose that the particle can freely move either to left side or right side, as shown in Figure 2, with a distance of IPS until it touches the nearest neighbors, then the free volume of this particle should be

$$V_{fp} = (2IPS)^3 = 64(\sqrt[3]{\phi_m / \phi} - 1)^3 r^3 \qquad (17$$

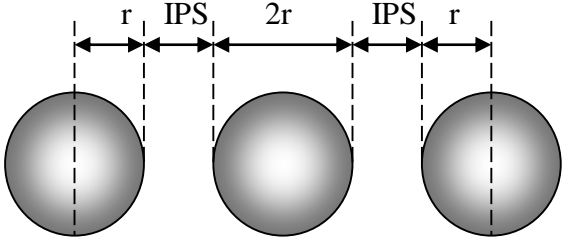

*Figure 2. Determination of the free volume of particles. From Hao, 2005, Electrorheological fluids: the non-aqueous suspensions, Amsterdam: Elsevier. With permission*

as the particle moves three dimensionally, where $V_{fp}$ is the free volume of an individual particle. The total free volume may be expressed as the free volume of an individual particle times the number of particles in the system,

$$V_f = 64(\sqrt[3]{\phi_m/\phi} - 1)^3 r^3 \times \frac{\phi V_s}{\frac{4\pi}{3}r^3}$$

$$= \frac{48\phi}{\pi}(\sqrt[3]{\phi_m/\phi} - 1)^3 V_s \tag{18}$$

Eq. (18) gives the total free volume when the particle volume fraction is $\phi$. Similarly when the particle volume fraction is $\phi_0$, the total free volume should be

$$V_{f0} = \frac{48\phi_0}{\pi}(\sqrt[3]{\phi_m/\phi_0} - 1)^3 V_s \tag{19}$$

Again, using the similar way to define the rate constant $k$ yields

$$k = A\frac{\phi(\sqrt[3]{\phi_m/\phi} - 1)^3}{\phi_0(\sqrt[3]{\phi_m/\phi_0} - 1)^3} \tag{20}$$

Eq. (9) may be analogically written as:

$$\frac{\phi_0(\sqrt[3]{\phi_m/\phi_0} - 1)^3}{15.29 A \phi^2(\sqrt[3]{\phi_m/\phi} - 1)^6} d\phi = \frac{dn}{n/n_m + 1} \tag{21}$$

Again, integrating both sides within the particle volume fraction range $(\phi_0, \phi)$ and using the boundary condition ($n=0$, the particle volume fraction is $\phi_0$, and $n=n$, the particle volume fraction is $\phi$) yields:

$$\frac{(\sqrt[3]{\phi_m}-\sqrt[3]{\phi_0})^3}{10}\left[\frac{(\phi_m^{2/3}-5\phi^{1/3}\phi_m^{1/3}+10\phi^{2/3})}{(\sqrt[3]{\phi_m}-\sqrt[3]{\phi})^5}-\frac{(\phi_m^{2/3}-5\phi_0^{1/3}\phi_m^{1/3}+10\phi_0^{2/3})}{(\sqrt[3]{\phi_m}-\sqrt[3]{\phi_0})^5}\right]=\frac{15.29A}{n_m}\ln\left(\frac{n}{n_m}+1\right) \quad (22)$$

For comparison, the particle volume fraction increase $(\phi-\phi_0)$ predicted with Eq. (11) and Eq. (22) is plotted against $(n+1)$ in Figure 3. It looks like the generalized Chicago logarithmic law expressed in Eq. (11) predicts a large increase of particle volume fraction under a small number of taps. In contrast, Eq. (22) gives a relative slow and gentle tapping process, though both two equations use same parameters ($A=2000$, $n_m=1000$, $\phi_m=0.63$) for calculation. The random dense packing structure with the maximum packing fraction 0.63 is assumed.

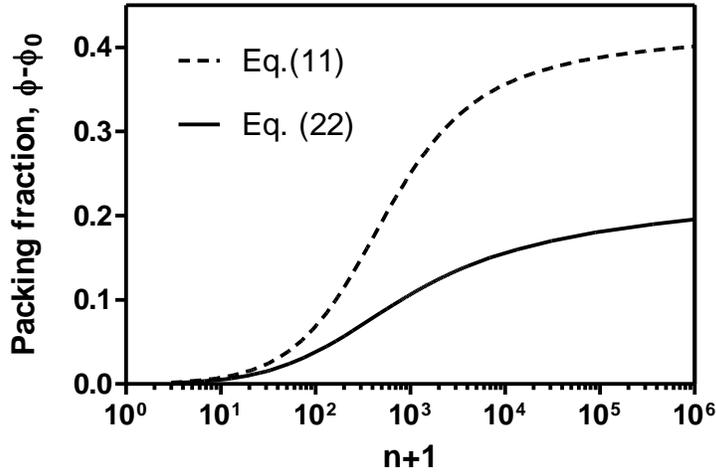

*Figure 3 Particle volume fraction increase $(\phi-\phi_0)$ predicted with Eq. (11) and Eq. (22) is plotted against (n+1) under assumptions A=2000, $n_m$=1000, $\phi_m$=0.63, the maximum particle packing fraction of random dense packing structure. The initial particle volume fraction is 0.2.*

Let's turn attention to the stretched exponential law. Suppose the powder is under a vertical tapping process, $L=L_0\exp(i\omega t)$, where $L_0$ is the tapping amplitude and $\omega$ is the tapping frequency. The energy flowing from the shaker to the powder (Hao 2014) is:

$$E(t)=\frac{MgL_0\omega}{\pi} \quad (23)$$

Where $Mg$ is the weight of powder. The total energy flowing to the powder after a series of tapping processes with the tapping number, $n$, may be expressed as:

$$E = n \bullet E(t) = \frac{nMgL_0\omega}{\pi} \qquad (24)$$

We may consider the tapping process as a rate process again and this rate process obeys the stretched Arrhenius equation as powder systems may be considered as disorder systems, in which many physical properties have the stretched exponential form, such as the conductivity of disordered systems that obeys Mott's variable range hopping model (Mott 1969). The stretched exponential function was first introduced by Kohlrausch in 1854 (Kohlrausch 1854) to describe the discharge phenomenon of a capacitor, later extended to describe dielectric spectra of polymers by Williams and Watts (Williams and Watts 1970), and now frequently applied to a large range of relaxations in disordered thermal systems such as glasses (Phillips 1996). The stretched Arrhenius equation may be expressed as:

$$k = A\exp\left[-\left(\frac{E}{RT}\right)^{\beta}\right] \qquad (25)$$

Where $k$ is the rate constant of a tapping process, and $\beta$ is the stretching of the exponential. Substituting Eq. (24) into Eq. (25) yields

$$k = A\exp\left[-\left(\frac{nMgL_0\omega}{\pi RT}\right)^{\beta}\right]$$
$$= A\exp\left[-\left(\frac{n}{\tau}\right)^{\beta}\right] \qquad (26)$$

Where $\tau$ is a constant of $\tau = \pi RT/(MgL_0\omega)$. We may use two approaches to represent the tapping rate constant. First, suppose that the tapping process rate constant is related to how the free volume of a granular system decreases with the number of taps, as shown in Eq. (6). Combining Eq. (6) and (26) together yields:

$$\frac{\phi_m - \phi}{\phi_m - \phi_0} = A\exp\left[-\left(\frac{n}{\tau}\right)^{\beta}\right] \qquad (27)$$

Re-arranging Eq. (27) yields:

$$\phi = \phi_m - A(\phi_m - \phi_0)\exp\left[-\left(\frac{n}{\tau}\right)^{\beta}\right] \qquad (28)$$

Eq. (28) is identical to Eq. (3) if *A* is assumed to be 1. Again, $\tau$ is a constant, $\tau = \pi RT/(MgL_0\omega)$, dependent on the tapping amplitude and frequency, the weight of the granular materials, and the temperature. Second, one may use Eq. (20) to express the tapping rate constant. Substituting *k* in Eq. (25) with Eq. (20) yields

$$\frac{\phi(\sqrt[3]{\phi_m/\phi}-1)^3}{\phi_0(\sqrt[3]{\phi_m/\phi_0}-1)^3} = A\exp\left[-\left(\frac{n}{\tau}\right)^\beta\right] \qquad (29)$$

Re-arrange Eq. (29) yields

$$\phi^{1/3} = \phi_m^{1/3} - A^{1/3}(\phi_m^{1/3} - \phi_0^{1/3})\exp\left[-\frac{1}{3}\left(\frac{n}{\tau}\right)^\beta\right] \qquad (30)$$

For comparing the difference between Eq. (28) and (30), the predicted packing fraction difference vs. the number of taps is plotted in Figure 4. At same relaxation time and initial packing fraction, Eq. (30) gives a relatively slower pace of compaction process with lower initial tapping impact on the packing fraction. For reaching a steady-state packing structure, Eq. (30) requires a large number of taps above $10^5$, while Eq. (28) predicts that the steady-state packing may be reached at relatively small number of taps about $10^3$. In both Eq. (28) and (30), there is one more parameter that may be used for fitting the data in comparison with the original stretched exponential law, Eq. (3), thus a better fit may be obtained with both Eq.(28) and (30) than Eq. (3).

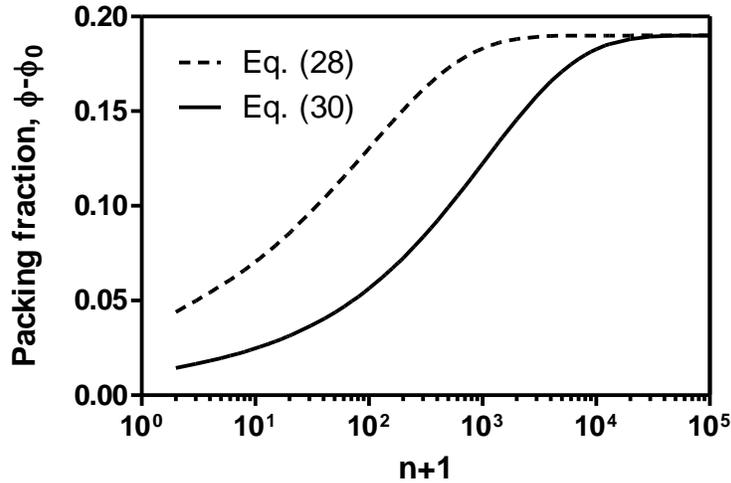

*Figure 4. The predicted packing fraction less the initial particle volume fraction with Eq. (28) and (30) is plotted again (n+1), n is the number of taps, under assumptions that the particles will have Rhombohedral pack structure with the maximum packing fraction 0.74, the initial particle packing fraction is 0.55, β=0.5, τ=100, and A=0.85.*

In summary, the rate process theory is utilized to treat the powder packing process and the free volume in a powder system is believed to control how quick particles can be packed. The derived equations can be reducible to the two widely used empirical equations under special circumstances, and could potentially fit the experimental data better with an additional adjustable parameter. The tapping amplitude and frequency, the weight of the granular materials, and the environment temperature are grouped into one parameter that weighs the pace of packing process. An exact same treatment method employing the theory of rate process and the free volume concept was successfully used to derive the viscosity equations of liquids, colloidal suspensions, and polymeric systems (Hao, 2005, Hao 2008), implying that both "dry" and "wet" particle systems are governed by the same physical mechanisms in term of the role of the free volume and how particles behave (a rate controlled process).

***Acknowledgments***: *The author appreciates Michelle Moyer and KaLee Dahlin for their generous support and constructive suggestions during the entire process of this work.*